\documentstyle[12pt,epsf]{article}

\begin{document}
\sloppy
\sloppy
\sloppy
$\ $
\begin{flushright}{UT-761, 1996\\hep-th/9610164} 
\end{flushright}
\vskip 0.5 truecm

\begin{center}
{\large{\bf  A Schwinger term in q-deformed su(2) algebra\footnote{To be published in Modern Physics Letters A}}}
\end{center}
\vskip .2 truecm
\centerline{\bf Kazuo Fujikawa and  Harunobu Kubo }
\vskip .2 truecm
\centerline {\it Department of Physics,University of Tokyo}
\centerline {\it Bunkyo-ku,Tokyo 113,Japan}
\vskip .2 truecm
\centerline{\bf C. H. Oh}
\vskip .2 truecm
\centerline {\it Department of Physics,National University of Singapore}
\centerline {\it Singapore 119260, Republic of Singapore}

\vskip 0.5 truecm


\begin{abstract}
An extra term generally appears in the q-deformed $su(2)$ algebra for the 
deformation parameter $q = \exp{ 2 \pi i\theta}$, if one combines the Biedenharn-Macfarlane construction of q-deformed $su(2)$, which is a generalization of Schwinger's construction of conventional $su(2)$, with the representation of the q-deformed
oscillator algebra which is manifestly free of negative norm.  This extra term introduced by the requirement of positive norm is analogous to the Schwinger term in current algebra. Implications of this extra term on the Bloch electron problem  analyzed by Wiegmann and Zabrodin are briefly discussed.

\par

\end{abstract}

The notion of q-deformed algebra[1], which was originally introduced in
connection with the inverse scattering problem and the Yang-Baxter
equation[2], is going to be a standard machinery of theoretical
physics. For example, the q-deformed $su(2)$ for $q=\exp{ i\pi P/Q}$
with mutually prime integers $P$ and $Q$ found a very interesting
physical application to the Bloch electrons  in two-dimensional lattice
model[3-5]. Also, the q-deformed oscillator algebra, which was
introduced by Biedenharn[6] and Macfarlane[7] to construct the
q-deformed $su(2)$ in the manner of  Schwinger's  construction of
conventional $su(2)$, found an interesting implication on the phase
operator problem of the photon[8-9]: The real-positive deformation
parameter $q$ or $q = \exp{2 \pi i\theta}$ with an irrational $\theta$
gives rise to the conventional Susskind-Glogower phase operator[10],
and $q =\exp{2 \pi i\theta}$ with a rational $\theta$ formally gives
rise to the Pegg-Barnett phase operator[11]. The singular nature of the
transition from one of  these two phase operators to the other by a
limiting procedure has been analyzed on the basis of the representation
of oscillator algebra[12]which is manifestly free of negative norm[13]
and the notion of index[14].

In some of  physical applications  of q-deformed algebra, the notion of
Hilbert space with positive definite norm is  crucial. This property of
positive norm is not quite transparent in the abstract mathematical
formulation of q-deformation. The purpose of the present paper is to
study to what extent the q-deformed $su(2)$ with  $q =\exp{2
i\pi\theta}$ is modified if one demands that the representation be
manifestly free from negative norm. (For real positive $q$, we do not
find an inevitable modification of algebra  on the basis of
positivity). Our basic strategy to study this problem is to start with
the Biedenharn-Macfarlane construction of $su(2)$ by using the
representation of the q-deformed oscillator algebra which is manifestly
free of negative norm. By this way, we can use the standard Fock space
technique with  positive definite norm. It is shown that we generally
find an extra term (``Schwinger term'') in the q-deformed algebra,
though in certain cases of physical interest this extra term
identically vanishes.

We start with the oscillator algebra introduced by Hong Yan [12]
\begin{eqnarray}
[a, a^{\dagger}] &=& [N_{a} +1] - [N_{a}]\nonumber\\
{[}N_{a}, a^{\dagger}{]} &=& a^{\dagger}\nonumber\\
{[}N_{a}, a {]} &=& -a\nonumber\\
c &=& a^{\dagger}a - [N_{a}]
\end{eqnarray}
and another set of oscillator variables $b, b^{\dagger}$ and $N_{b}$.
The value of the Casimir operator $c$ is chosen to be identical for
these two sets of oscillators: $c = b^{\dagger}b - [N_{b}]$. The usual
notation of $[X] = \sin( 2\pi\theta X)/\sin(2\pi\theta)$ for the
deformation parameter $q=\exp{2\pi i\theta}$ with $-1/2 < \theta < 1/2$
is used.  It is known that the algebra (1) supports the Hopf
structure[12][15] but not the q-oscillators employed by Refs.[6,7].
Furthermore the latter q-oscillators suffer from a negative norm
problem when $q = \exp 2 \pi i \theta$ for generic $\theta$.

The representation of the oscillator algebra (1) free of negative norm
is defined by[13]
\begin{eqnarray}
|l\rangle_{a} &=& \frac{1}{\sqrt{([l-n_{0}] + [n_{0}])!}}
(a^{\dagger})^{l}|0\rangle\nonumber\\ |l\rangle_{b} &=&
\frac{1}{\sqrt{([l-n_{0}] + [n_{0}])!}} (b^{\dagger})^{l}|0\rangle
\end{eqnarray}
with $l= 0,1,2,...$ and the number $n_{0}$, which characterizes the
Casimir operator  $c$, is defined to satisfy
\begin{equation}
c = [n_{0}] = \frac{\sin 2\pi n_{0}\theta}{\sin 2\pi \theta}=
\frac{1}{|\sin 2\pi\theta |}
\end{equation}
 for $\theta \neq 0$. We also set $a|0\rangle = b|0\rangle = 0$. We
then have
\begin{eqnarray}
a|l\rangle_{a} &=& \sqrt{[l-n_{0}] + [n_{0}]}|l-1\rangle_{a}\nonumber\\
a^{\dagger}|l\rangle_{a}&=& \sqrt{[l+1 -n_{0}] +
[n_{0}]}|l+1\rangle_{a}
\nonumber\\
N_{a}|l\rangle_{a}& =& (l-n_{0})|l\rangle_{a}
\end{eqnarray}
and similarly for $|l\rangle_{b}$.

It is obvious that $[l] = \sin 2\pi\theta l{/}\sin 2\pi\theta$ can be
negative as well as positive for $\theta \neq 0$. In contrast, for the
choice of the Casimir operator in (3), we can confirm
\begin{eqnarray}
[l-n_{0}] + [n_{0}] &=& ( -\cos 2\pi\theta l + 1)\frac{\sin 2\pi\theta
n_{0}} {\sin 2\pi\theta}\nonumber\\ &=& \frac{1}{|\sin 2\pi\theta
|}(1-\cos 2\pi\theta l) \geq 0
\end{eqnarray}
and thus the representation (4) is free of negative norm for an
irrational $\theta$. We thus have
\begin{equation}
\langle l| l^{\prime}\rangle_{a} = \delta_{l l^{\prime}}
\end{equation}
if $\langle 0| 0\rangle = 1$, and $(a)^{\dagger} = a^{\dagger}$;
similar relations hold for $b$ operators. For a rational $\theta =
M/L$, the representation (2) is truncated at $l=L-1$ but still free of
negative norm. Our choice of $c=[n_{0}]$ in (3) ensures the absence of
negative norm. Conversely, one can confirm that $[l-n_{0}] + [n_{0}]$
in (5) is made arbitrarily close to zero for a suitable $l (\neq 0)$
for any given $\theta (\neq 0)$, by noting that $\theta$ can be
approximated arbitrarily accurately by a rational number (i.e., by a
ratio of  sufficiently large integers), though this does not
necessarily mean that the transition from a rational number to an
irrational one is smooth. In this sense, our modification of the
representation by the Casimir operator $c$ is {\em minimal}. This
minimal property becomes important later, since the presence of the
``Schwinger term'' in q-deformed $su(2)$ to be defined later then
suggests the inevitable presence of {\em some} representations which
are inflicted by negative norm, if one sets $c=[n_{0}] = 0$ there.

On the basis of the representations (2) and (3), we define the
Biedenharn-Macfarlane construction of q-deformed $su(2)$ generators by
\begin{eqnarray}
S_{+} &=& a^{\dagger}b\nonumber\\ S_{-} &=& b^{\dagger}a\nonumber\\
S_{3} &=& \frac{1}{2}(N_{a} - N_{b})\nonumber\\ {\cal C} &=&
\frac{1}{2}(N_{a} + N_{b})
\end{eqnarray}
where ${\cal C}$ stands for the Casimir operator of this algebra. On
the basis of this definition one finds
\begin{eqnarray}
{[}S_{\pm}, S_{3}{]} &=& \mp S_{\pm}\nonumber\\ {[}S_{+}, S_{-}{]}&=&
[2S_{3}] + c \{ [N_{b}+1] -[N_{b}] - [N_{a} + 1] + [N_{a}]\}\nonumber\\
&=& [2S_{3}] + 4[n_{0}]\sin \pi\theta \sin 2\pi\theta [S_{3}][{\cal C}
+ \frac{1}{2}]
\end{eqnarray}
The last term in (8), which is proportional to the Casimir operator $c$
of the oscillator algebra in (3), gives rise to an extra term in the
conventional q-deformed $su(2)$ algebra. This extra term emerges
through the use of the q-oscillator (1) which has a positive norm
representation.  The basic reasoning for the existence of the
conventional Schwinger term in current algebra [16] was the energy
spectrum bounded from below and the positive norm of the Hilbert space.
The present construction of (7) may be regarded as a simplest version
of current algebra, and for this reason we tentatively call this extra
term in (8) as ``Schwinger term'', though a more suitable terminology
for it may exist. We note that the modified algebra in (1) with a
non-trivial Casimir operator, instead of the oscillator algebra in
Refs.[6-7] which is obtained by setting $c=0$ in (1), is crucial to
ensure the absence of the negative norm. One can define (8) for a
general value of $n_{0}$, but to ensure the absence of the negative
norm one has to choose $n_{0}$ as in (3), which implies that $n_{0}$ is
a function of $\theta$. Because of this property, the Schwinger term
does not vanish in general even in the limit $\theta \rightarrow 0$..
The limit $\theta \rightarrow 0$ is generally singular as was
emphasized in Ref.[13].

A $d=2j+1$ dimensional (highest weight) representation of the algebra
(8) with a Schwinger term is defined on the oscillator Fock space in
(2) by
\begin{eqnarray}
S_{+}|j,m\rangle& =& \sqrt{([j+m+1-n_{0}] + [n_{0}])([j-m-n_{0}] +
[n_{0}])} |j,m+1\rangle\nonumber\\ S_{-}|j,m\rangle &=&
\sqrt{([j-m+1-n_{0}] + [n_{0}])([j+m-n_{0}] + [n_{0}])}
|j,m-1\rangle\nonumber\\ S_{3}|j,m\rangle &=& m|j,m\rangle
\end{eqnarray}
where
\begin{equation}
|j,m\rangle = |j+m\rangle_{a}\otimes |j-m\rangle_{b}, \ \ \ m= -j,
-j+1, ...,j
\end{equation}
with $j= 0, 1/2, 1, 3/2,......$, and the orthonormality relation
\begin{equation}
\langle j,m|j^{\prime},m^{\prime}\rangle = \delta_{j
j^{\prime}}\delta_{m m^{\prime}}
\end{equation}
Note that we can satisfy the basic requirement
\begin{equation}
(S_{+})^{\dagger} = S_{-}
\end{equation}
for the representation in (9), and the highest weight condition
$S_{+}|j,j\rangle = S_{-}|j,-j\rangle = 0$.

The $2j+1$ dimensional highest weight representation of the algebra (8)
can also be realized by $q-$difference equations as
\begin{eqnarray}
\tilde{S}_{+}\psi(z)&=&(q-q^{-1})^{-1}z(q^{2j-n_{0}}\psi(q^{-1}z)-q^{-2j+n_{0}}\psi(qz))
+z[n_0]\psi(z), \nonumber \\
\tilde{S}_{-}\psi(z)&=&-(q-q^{-1})^{-1}z^{-1}(q^{n_{0}}\psi(q^{-1}z)-q^{-n_{0}}\psi(qz))
+z^{-1}[n_0]\psi(z),  \\ q^{\tilde{S}_{3}}\psi(z)&=&q^{-j}\psi(qz),
\nonumber
\label{def}
\end{eqnarray}
where $\psi(z)$ is a polynomial of degree $2j$.  This representation
satisfies the highest weight condition $\tilde{S}_{+}z^{2j}=0$ and the
lowest weight condition $\tilde{S}_{-}\cdot 1 =0$. The representation
of (13) for the bases, $z^{j+m},\ m= j, j-1, ......, -j$, is given by
\begin{eqnarray}
\tilde{S}_{+}z^{j+m} &=& ([j - m - n_{0}] + [n_{0}])z^{j+m+1}\nonumber\\
\tilde{S}_{-}z^{j+m} &=& ([j + m - n_{0}] + [n_{0}])z^{j+m-1}\nonumber\\
q^{\tilde{S}_{3}}z^{j+m} &=& q^{m}z^{j+m}
\end{eqnarray}
This representation is related to the standard representation in (9) by
a (diagonal) similarity transformation $A$; \ $\tilde{S}_{+} = A S_{+}
A^{-1}$ and $\tilde{S}_{-} = A S_{-} A^{-1}$. Note that $A$ is not
unitary, and $\tilde{S}_{+}^{\dagger} \neq \tilde{S}_{-}$.

We now discuss the possible implications of our representation (9). A
very specific $d=2j+1$ dimensional representation for the value of the
deformation parameter $q=e^{2\pi i\theta}$ where
\begin{equation}
\theta = \frac{P}{2Q} = \frac{P}{2(2j+1)}
\end{equation}
with mutually prime integers $P$ and $Q$ found an interesting
application in the Bloch electron problem[3-5]. Note that $Q$ and the
dimension of the representaion $d=2j+1$ are independent in general, but
in the present case they are related in a specific way. Our states in
(2) are sufficient to support this representation since the states in
(2) for the value of $\theta$ in (15) form a $(2j+1)$-dimension space
for $P=$ even and a $2(2j+1)$-dimensional one for $P=$odd. For the
value of $\theta$ in (15), the Schwinger term in (8) becomes by noting
$2{\cal C} + 1 = 2j+1-2n_{0}$ for a $2j+1$ dimensional representation,
\begin{eqnarray}
4\sin \pi\theta (\frac{\sin 2\pi\theta }{|\sin 2\pi\theta|})[S_{3}]
\frac{\sin\pi\theta (2j+1-2n_{0})}{\sin 2\pi\theta}&=&
\frac{-2}{\cos\pi\theta}\cos \pi \theta (2j + 1) [S_{3}]\nonumber\\
&=& \frac{-2}{\cos \pi\theta}\cos (\frac{\pi}{2}P)[S_{3}]
\end{eqnarray}
where we have used $\sin 2\pi n_{0}\theta = \sin 2\pi\theta /|\sin
2\pi\theta |,\ \
\sin^{2} 2\pi n_{0}\theta = 1$, and $\cos 2\pi n_{0}\theta = 0$. 
( Note that the case $j=0$ and $P=$odd is excluded here due to the
constraint $-1/2 < \theta < 1/2$ to define $[X] = \sin (2\pi\theta
X)/\sin (2\pi\theta )$ for general $X$.) The Schwinger term in (16)
identically vanishes for $P=${\em  odd}, which is one of the allowed
cases  in the analysis in Ref.[3] and the case analyzed in great detail
in Ref.[4]. For this specific case, the Schwinger term identically
vanishes and the conventional representation of q-deformed $su(2)$
becomes free of negative norm. In fact, for $P=$ odd, one can confirm
that our representation in (9) with the value of $n_{0}$ specified
there is precisely re-written as
\begin{equation}
S_{+}|j,m\rangle = \sqrt{[j+m+1][j-m]} |j,m+1\rangle
\end{equation}
and $S_{-} = (S_{+})^{\dagger}$.  This (17) is the conventional
representation with $c=[n_{0}]=0$.

On the other hand, for  $P=$ even, the Schwinger term does not vanish.
This suggests that the conventional representation of q-deformed
$su(2)$ with $c=[n_{0}]=0$ in (9) generally contains  negative norm.
This is in fact confirmed by noting that
\begin{eqnarray}
[j+m+1][j-m] &=& \frac{1}{\sin^{2} 2\pi\theta }\sin 2\pi\theta
(j+m+1)\sin 2\pi\theta (j-m)\nonumber\\ &=&
\frac{1}{2\sin^{2}(\pi\frac{P}{2j+1})}\{\cos (\pi P\frac{2m+1}{2j+1})
\pm 1\}\end{eqnarray} for $P=$ odd and $P=$ even, respectively.  The
minus sign in (18) holds for $P=$ even, and $[j+m-1][j-m]$ becomes {\em
non-positive} , which spoils $S_{-} = (S_{+})^{\dagger}$ and induces
negative norm into the Fock space if one chooses $ c= [n_{0}] = 0$ in
(9);  in fact, one has $S_{-} = - (S_{+})^{\dagger}$ for  $ c= [n_{0}]
= 0$. From this view point, it is seen why the {\em hermitian}
Hamiltonian in Ref.[3], where the case $ c= [n_{0}] = 0$ is considered,
is fitted by
\begin{equation}
H = i (q - q^{-1}) (S_{-} \pm S_{+})
\end{equation}
with $\pm$ sign corresponding to $P=$ odd(even), respectively. To be
precise,  $$H = i(q - q^{-1})(\rho(S_{-}) ~ \pm ~ \rho(S_{+}))$$
by using the
cyclic representation in eq(20) below.  A
cyclic representation corresponding to (9) is  obtained by putting
$z=q^{k}, (k=1,2,\cdots,2Q)$ in (13) for the value of $\theta$ in (15).
There are $2Q$ bases $\psi_{k}\equiv \psi(q^{k})$ which satisfy
$\psi_{k+2Q}=\psi_{k}$, and we define
\begin{eqnarray}
\rho (S_{+})\psi_{k}&=&\pm (q-q^{-1})^{-1}(q^{k+1+n_{0}}\psi_{k+1}
                                     -q^{k-1-n_{0}}\psi_{k-1})
                 +q^{k}[n_0]\psi_{k}  \nonumber\\
\rho (S_{-})\psi_{k}&=&(q-q^{-1})^{-1}(q^{-k-n_{0}}\psi_{k+1}
                                     -q^{-k+n_{0}}\psi_{k-1})
                 +q^{-k}[n_0]\psi_{k} \nonumber\\
q^{\rho (S_{3})}\psi_{k}&=&q^{-j}\psi_{k+1} 
\end{eqnarray}
where $\pm$ sign corresponds to $P=$ odd(even), respectively.
It is confirmed 
that this cyclic representation $\rho(S)$ satisfies the algebra (9)
with a Schwinger term given by (16). In particular, the Schwinger term
vanishes for $P=$ odd if one notes $\cos 2\pi n_{0}\theta = 0$. This
means that the cyclic representation (20) for $P=$ odd is equivalent to
the conventional one in Ref.[3] with $ c= [n_{0}] = 0$. A physical
significance of the representation (20) for $P=$ even with the
Schwinger term in (16) is yet to be seen: Group theoretically, one
could use $H = i (q - q^{-1}) (S_{-} + S_{+})$ in (19) even for $P=$
even if $ c= [n_{0}]$ is chosen as in (3).

The Schwinger term in (8) vanishes for $\theta = \frac{P}{2(2j+1)}$
with $P=$ odd, whereas the Schwinger term is required to preserve the
positive norm of the Hilbert space for $\theta = \frac{P}{2(2j+1)}$
with $P=$ even or an irrational $\theta$. This fact might be related to
the findings in Ref.[4]; it is shown there that the definition of the
case of an irrational $\theta$ as a limiting case of $\theta =
\frac{P}{2(2j+1)}$ with odd $P$ leads to a singular (not differentiable
anywhere) behavior of a certain quantity in the Bloch electron problem.

In conclusion, we generally find a Schwinger term in the q-deformed
$su(2)$ algebra for $q=e^{2\pi i \theta}$, if one follows the
Biedenharn-Macfarlane construction on the basis of the oscillator
algebra representation which is manifestly free of negative norm.
Mathematically it is not known at this moment if the modification of the 
$q$-deformed $su(2)$ algebra by the Schwinger term preserves the Hopf 
structure or not, but we believe that
it is sensible to impose the positive
definite norm on the Fock space and to see its physical implications.
At least, our Schwinger term is a simple and reliable indicator of
negative norm for the  representation with $q=e^{2\pi i \theta}$: If
the Schwinger term vanishes for a specific representation, it
definitely shows that the corresponding conventional representation
with $c=[n_{0}]=0$ is free of negative norm. On the other hand, the
presence of the Schwinger term shows the existence of {\em some}
representations which are inflicted with negative norm if one sets
$c=[n_{0}]=0$ in (9).

\end{document}